**High quality epitaxial $FeSe_{0.5}Te_{0.5}$ thin films grown on $SrTiO_3$ substrates by pulsed laser deposition**

E.Bellingeri[a], R.Buzio[a], A.Gerbi[a], D.Marrè[a,c], S.Congiu[a], M.R.Cimberle[b], M.Tropeano[a,b], A.S.Siri[a,c], A. Palenzona[a,d] C.Ferdeghini[a],

[a] *CNR/INFM-LAMIA corso Perrone 24, 16152 Genova, Italy*

[b] *CNR-IMEM, Dipartimento di Fisica, Via Dodecaneso 33, 16146 Genova, Italy*

[c] *Dipartimento di Fisica, Via Dodecaneso 33, 16146 Genova, Italy*

[d] *Dipartimento di Chimica e Chimica Industriale, Via Dodecaneso 31, 16146 Genova, Italy*

E-mail: emilio.bellingeri@lamia.infm.it

**Abstract**

Superconducting epitaxial $FeSe_{0.5}Te_{0.5}$ thin films were prepared on $SrTiO_3$ (001) substrates by pulsed laser deposition. The high purity of the phase, the quality of the growth and the epitaxy were studied with different experimental techniques: X-rays diffraction, reflection high energy electron diffraction, scanning tunnelling microscopy and atomic force microscopy. The substrate temperature during the deposition was found to be the main parameter governing sample morphology and superconducting critical temperature. Films obtained in the optimal conditions show an epitaxial growth with c axis perpendicular to the film surface and the a and b axis parallel to the substrates one, without the evidence of any other orientation. Moreover, such films show a metallic behavior over the whole measured temperature range and critical temperature above 17K, which is higher than the target one.

**Introduction**

Since the discovery of superconductivity in the compound $LaFeAsO_{(1-x)}F_x$ [1] an enormous effort in studying this type of materials and related compounds has brought to the discovery of new families of superconductors: in particular K or Na doped $AFe_2As_2$ (122 phase with A= Ba,Sr,Ca), (Li,Na)FeAs (111 phase) and, more recently, FeSe (11 phase) ( for a review see [1] and references therein). The study of the intrinsic properties of these materials needs that single crystals or epitaxial thin films may be available. On the other side thin films are also requested for potential technological applications. Few studies on superconducting thin films of the 111 and 122 phases have been reported, but a very small work has been done so far on the deposition of thin films of the 11 phase. The latter case appears of great interest for several reasons: first of all it would be easier the deposition of a material with only two components with respect to other multicomponents phases; secondly, the lack of As avoids all the problems related to the management of a poisoning element. Moreover, since in bulk materials it has been observed that the partial substitution of Se with Te produces a significant critical temperature increase [2] (optimal Tc for Se/Te=0.5/0.5), and the tetragonal phase is stabilized in a larger composition range [3] -making less critical the synthesis conditions- it seems appealing to grow $FeSe_{0.5}Te_{0.5}$ films.

It is finally well established that the external pressure increases the critical temperature of these compounds: from 8-10K up to about 37 K with the application of 8 GPa in the case of FeSe [4] and from 14 K to 26K at 2 GPa for $FeSe_{0.5}Te_{0.5}$[5]. As a consequence it is intriguing to hypothesise that strain effects due to the substrate may act in thin films as the external pressure in bulk and may produce significant enhancements in the critical temperature.

In the last ten years FeSe thin films for spintronic application have been grown on semiconductor substrates by selenization process, but no superconductivity was observed [6,7]. Recently, after the discovery of superconductivity in this system, three groups have produced superconducting FeSe thin films by laser ablation technique [8-10]. Only one group reported preliminary results on the deposition of $FeSe_{0.5}Te_{0.5}$ thin films [11]. Here we present the growth of fully epitaxial thin films of $FeSe_{0.5}Te_{0.5}$ on $SrTiO_3$ substrates by pulsed laser deposition (PLD) with optimal critical temperature. We report also their structural, morphological and physical characterization.

## Experimental and experimental results

A $FeSe_{0.5}Te_{0.5}$ bulk compound was prepared by direct synthesis from high purity materials (Fe 99.9+%, Se 99.9% and Te 99.999%) with a two steps procedure. Firstly a mixture of the starting elements and precursors has been reacted in a pyrex tube at 400°-450° C for 15-20 hours, than these first products were grinded, pellettized and heated at 800° C in a SiO2 tube for 7-8 days. All the operations were carried out in a glove box where $O_2$ and $H_2O$ were less than 1 ppm.

X-ray diffraction showed generally almost single phase products, sometimes with very small quantity of extra phases, 2-3 % at most. The bulk showed a sharp resistive transition with an onset (90% of the normal state) at T=16.2K, which is higher in respect to those reported in the current literature for Fe(Se,Te) compounds [2]. Magnetic characterization, performed with a SQUID magnetometer (Quantum Design), showed a nearly complete shielding of 90% at a field of 1mT, confirming the good quality of the target.

The films were deposited on single crystal strontium titanate (STO) (001) substrates in a ultra high vacuum PLD system using the above described bulk as a target [12]. The substrates were glued with silver paint onto a stainless steel sample holder compatible with the used Scanning Tunnel Microscope (STM) (Omicron Nanotechnology GmbH). Deposition temperature, as measured by an infrared pyrometer, ranged between 300 and 650 °C. The films were deposited under high vacuum conditions at a residual gas pressure of 5 $10^{-9}$ mbar at the deposition temperature. The quality of the growth was *in-situ* monitored by Reflection High Energy Electron Diffraction (RHEED) analysis. The laser beam (KrF 248 nm) was focalized onto a 2 $mm^2$ spot on the target with a fluency of 2 $J/cm^2$. Repetition rates ranging from 3 to 10 pulse/s were adopted and the target-substrate distance was kept fixed at 5 cm.

The deposited films were transferred at a pressure of $10^{-7}$ mbar to the STM system for surface characterization. Successively they were characterized *ex situ* by Atomic Force Microscopy (AFM) and X-ray diffraction (XRD). The electrical and superconducting properties were measured by a standard four wire technique in a home made cryostat employing a calibrated Cernox thermometer for a precise determination of the transition temperature.

XRD analysis showed that the $FeSe_{0.5}Te_{0.5}$ PbO-like tetragonal phase was found stable for all the explored deposition conditions; no traces of elementary oxides or of the hexagonal phase were observed even at the highest temperatures, in contrast to what observed in [11]. In the θ-2θ scans, as shown in figure 1, only the (00l) reflections of the film and substrate are present indicating the excellent purity of the phase and the optimum c-axis alignment of the growth. In figure 1a a detailed scan of the 001 reflection of $FeSe_{0.5}Te_{0.5}$ clearly shows finite size effect, indicating the perfect crystalline structure of the deposited layer. Film thickness was estimated both from reflectivity and finite size effect and values ranging from 17 nm to 90 nm were observed. Surprisingly the deposition rate was strongly affected by the deposition temperature: at the highest deposition temperatures used the rate was 0.01 Å *per* laser shot whereas for temperatures below 500 °C a rate of 0.04 Å *per* laser shot was observed.

In figure 2, $\phi$ scans of the (101) reflection of the film ($2\theta$=28.13°; $\chi$=33°, lower panel) and of the substrate ($2\theta$=32.16°; $\chi$=45°, upper panel) indicated the epitaxial growth of the $FeSe_{0.5}Te_{0.5}$, with the a and b axis parallel to the substrate ones without the evidence of any other orientation. The high quality of the growth and the epitaxy with the substrate was also confirmed by the RHEED images obtained for different in plane angles of the e-beam respect to the substrate reported in figure 2a. The position in circles of elongated spot in the RHEED pattern indicates a *quasi* 2D kind of growth as, also confirmed by STM and AFM measurements.

The surface roughness of thin films, estimated by AFM under ambient conditions, was in the range $0.7\text{nm} - 4\text{nm}$ depending on the chosen values of growth temperature and deposition rate. In particular at 400°C – 520°C and 3Hz respectively, we observed the presence of tiny isolated dots which a surface density of 500 – 700 $\mu m^{-2}$ and maximum height below 4nm. On increasing the deposition temperature and repetition rate, dots density decreased whereas their height significantly enlarged, to achieve values around 200 $\mu m^{-2}$ and 18nm respectively at 600°C and 5Hz. Since XRD analysis attests the presence of a single phase for our specimens, we argue that dots indeed preserve the same chemical composition and crystalline structure of the superconducting phase; their presence should be therefore ascribed to the specific mechanisms governing the growth process. Moreover we note that terraces with monoatomic steps of 0.5 – 0.6 nm in size were clearly resolved in STM images acquired at room temperature, confirming the good quality of depositions (Figure 3).

The deposition temperature was found to be the main parameter governing both the resistivity versus temperature behavior and the superconducting critical temperature of the deposited samples. Indeed, as shown in Figure 4c), the transition temperature raised with the deposition temperature up to 450°C, where the maximum of $T_c$ was found. Then, the critical temperature decreased and abruptly vanished for samples deposited above 600°C. It is worth noting that the best films had a critical temperature of 17K, which is higher than the target one.

Also the normal state resistivity behavior was influenced by the deposition temperature, as seen in figure 4a) showing the temperature dependence of the resistivity of selected films deposited at different temperatures. For comparison, also the resistivity of the target is reported. A nearly metallic behavior over the whole measured temperature range was found for the specimens deposited around 450°C, having the maximum $T_c$ values : on the contrary, on increasing or decreasing the growth temperature $T_c$ was reduced and the resistivity curves showed semiconducting features.

In summary we succeeded in depositing high quality $FeSe_{0.5}Te_{0.5}$ thin films. These films showed a very good epitaxy and an appreciable surface morphology. Moreover they revealed metallic behavior with a $T_c$ up to 17K, which higher than the critical temperature of the bulk target. One of the possible explanations for this phenomenon is that strain effects, induced by the substrate, may act in thin films as the external pressure in bulk, thus leading to a significant enhancement of the critical temperature. An investigation concerning the growth of $FeSe_{0.5}Te_{0.5}$ thin films on different substrates is in progress.

**Figure captions**

Figure 1. XRD θ-2θ scan of Fe(Se,Te) film deposited on SrTiO3. Only the 00*l* reflections of the film and substrate are detectable. In b) a detailed scan of the 001 reflection of the film showing finite size effect is reported.

Figure 2. a) ϕ scan of the (101) reflection of the film (lower panel) and of the substrate (upper panel) indicating the epitaxial growth of the Fe(Se,Te) with the a and b axis parallel to the substrates ones.
b) RHEED images acquired at the end of the growth for different in-plane angles also showing the epitaxiality of the growth: from the left to the right: 45°, 0° and 0°< θ<45°.

Figure 3. a) AFM topography acquired on a $FeSe_{0.5}Te_{0.5}$ thin film deposited at 400°C; scan area $500 \times 500 nm^2$. The inset shows terraces resolved by STM on a $58 \times 58 nm^2$ region of a film deposited at 450°C; voltage bias 1V, tunnelling current 1nA. b) Scan line corresponding the inset of panel a): atomic steps of 0.5nm-0.6nm in size are resolved.

Figure 4 a) resistivity as a function of temperature for selected thin films and the target: a change from a semiconducting to metallic behavior is observed depending on the deposition temperature (indicated in the legend). b) Magnification of the superconducting transition region. c) Critical temperature versus deposition temperature; as reference, the dashed line indicates the target critical temperature.

Figure 1

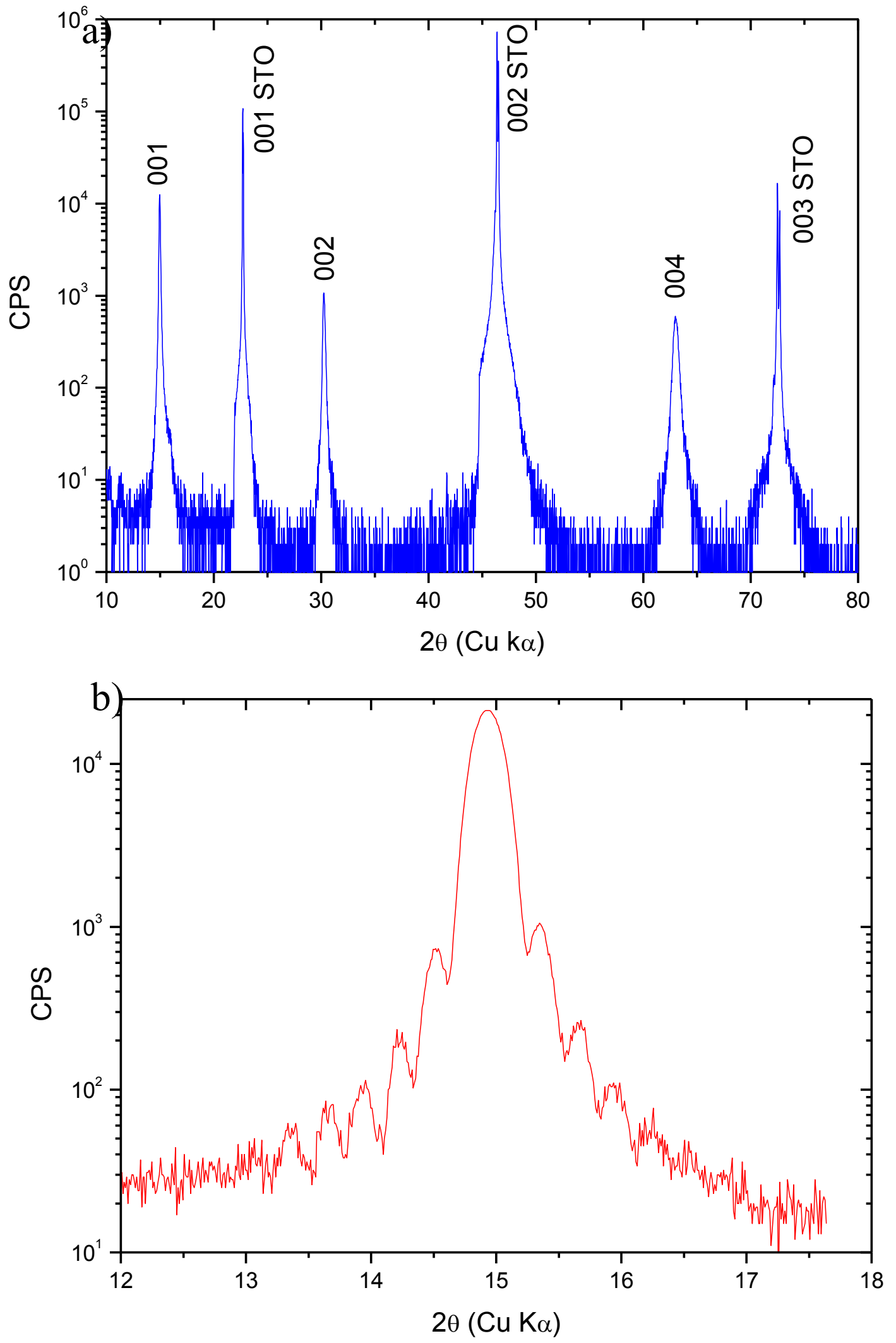

Figure 2

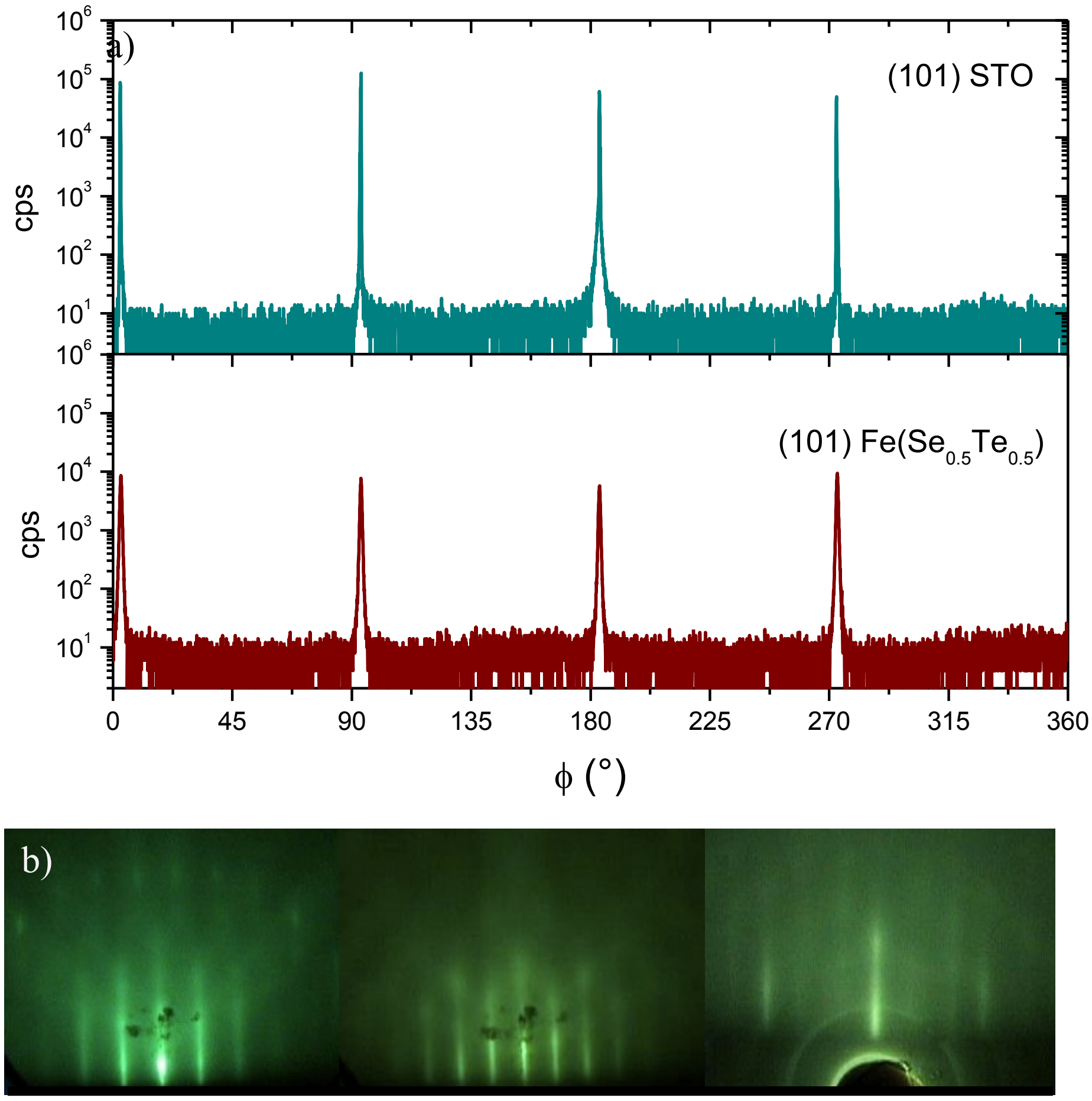

Figure 3

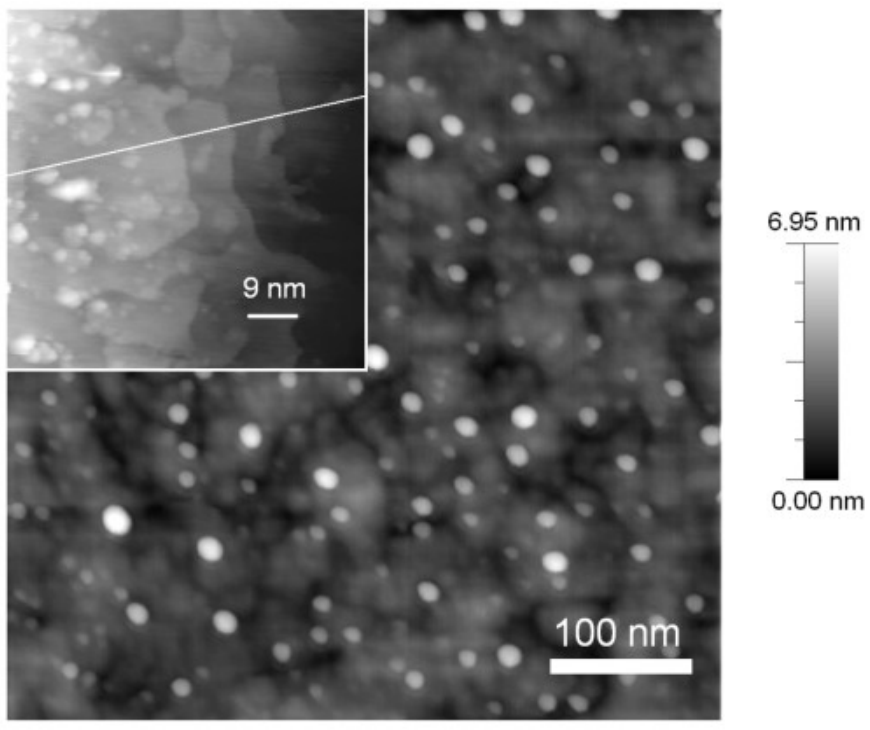

a)

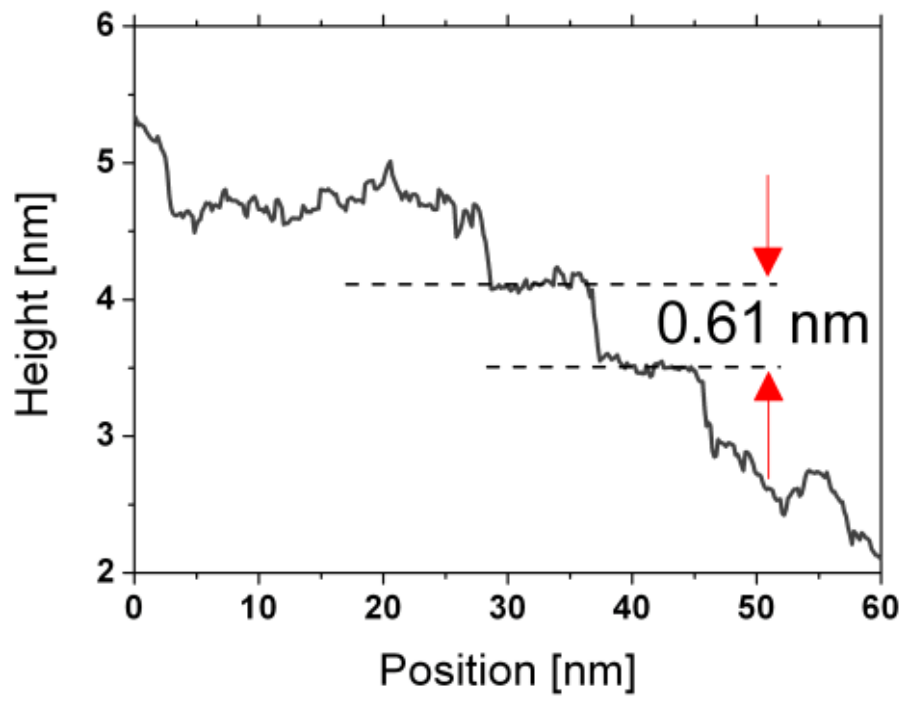

b)

Figure 4

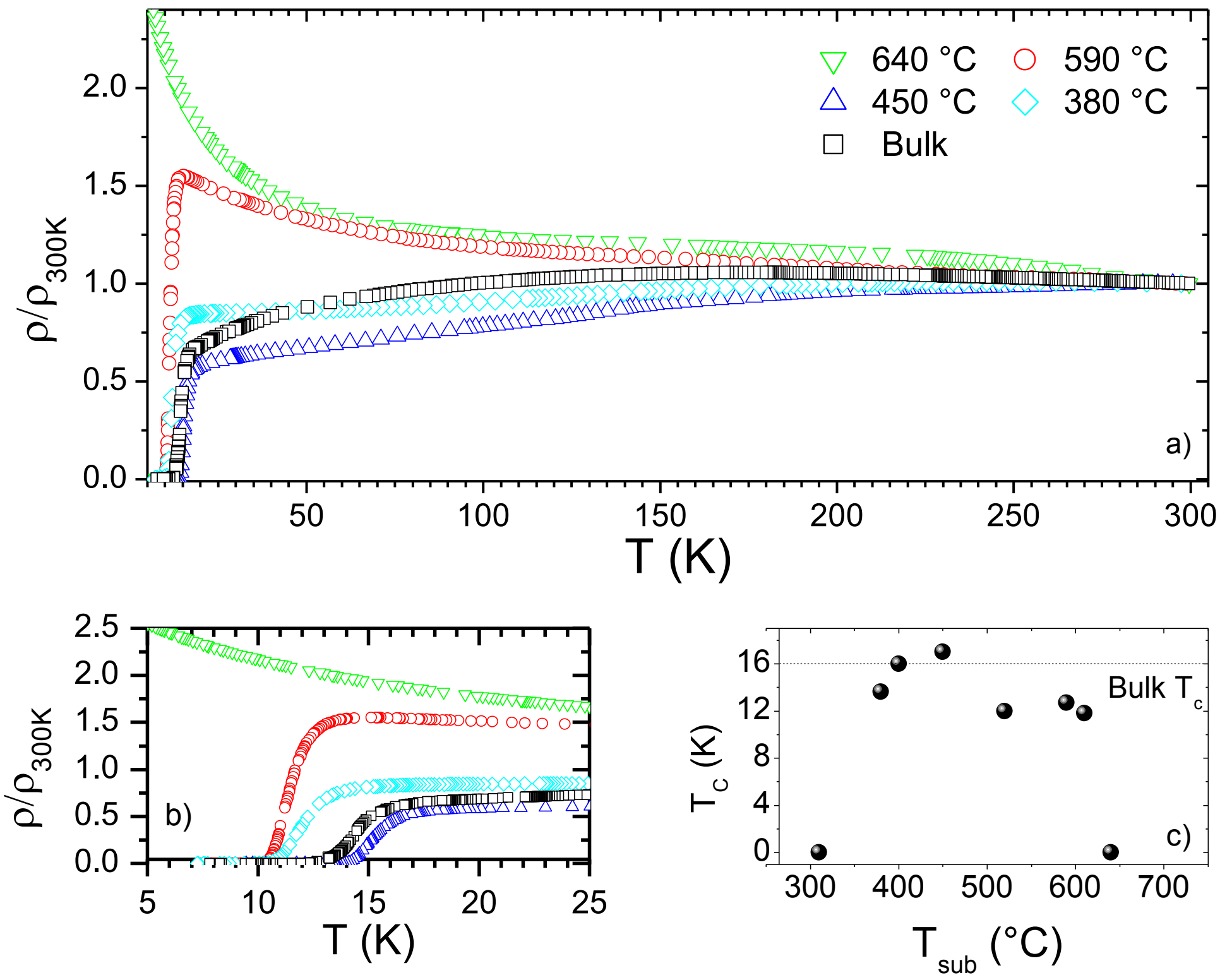